\documentclass[conference]{IEEEtran} 
\usepackage{cite}
\usepackage{lipsum} 
\usepackage{array} 
\usepackage{booktabs} 
\usepackage{caption} 
\usepackage{graphicx}

\title{LLMs Working in Harmony: A Survey on the Technological Aspects of Building Effective LLM-Based Multi Agent Systems}

\author{
    \IEEEauthorblockN{RM Aratchige}
    \IEEEauthorblockA{
        Department of Computer Science,\\
        Faculty of Computing,\\
        General Sir John Kotelawala Defence University,\\
        Ratmalana, Sri Lanka\\
        39-bse-0006@kdu.ac.lk
    }
    \and
    \IEEEauthorblockN{Dr. WMKS Ilmini}
    \IEEEauthorblockA{
        Department of Computer Science,\\
        Faculty of Computing,\\
        General Sir John Kotelawala Defence University,\\
        Ratmalana, Sri Lanka\\
        kalaniilmini@kdu.ac.lk
    }
}

\begin{document}

\maketitle

\begin{abstract}
This survey investigates foundational technologies essential for developing effective Large Language Model (LLM)-based multi-agent systems. Aiming to answer how best to optimize these systems for collaborative, dynamic environments, we focus on four critical areas: Architecture, Memory, Planning, and Technologies/Frameworks. By analyzing recent advancements and their limitations—such as scalability, real-time response challenges, and agent coordination constraints—we provide a detailed view of the technological landscape. Frameworks like the Mixture of Agents architecture and the ReAct planning model exemplify current innovations, showcasing improvements in role assignment and decision-making. This review synthesizes key strengths and persistent challenges, offering practical recommendations to enhance system scalability, agent collaboration, and adaptability. Our findings provide a roadmap for future research, supporting the creation of robust, efficient multi-agent systems that advance both individual agent performance and collective system resilience.
\end{abstract}


\section{Introduction}

\textit{“Individually, we are one drop. Together, we are an ocean.”}
\hspace*{\fill}- Ryunosuke Satoro


The advent of Large Language Models (LLMs) has transformed artificial intelligence, with the introduction of the Transformer architecture in the landmark paper “Attention is All You Need” [1] marking a key turning point. The Transformer replaced traditional sequence models like recurrent neural networks with an attention-based mechanism, boosting machine translation performance and reducing training time. Since then, LLMs have evolved further, particularly with models like GPT, which demonstrate unprecedented performance in natural language processing tasks [2]. LLMs now handle tasks ranging from text generation to summarization, enabling software that can understand and reason with natural language. This versatility drives the current research: exploring how LLMs can build complex multi-agent systems that collaborate and specialize in tasks, especially in environments demanding more than a single model’s capabilities.

Despite their successes, LLM-based applications have limitations. Hallucination remains a major issue, as models can produce inaccurate information without external validation, as noted by Adewumi et al. [3]. This limits reliability where precision is essential. LLMs also struggle with complex or abstract concepts, a challenge discussed by Cherkassky et al. [4], who showed that even advanced models like GPT-4 often fall short in imitating human reasoning. Multi-agent systems can mitigate these issues, allowing distinct agents to collaborate on complex tasks, as explored by Han et al. [5]. Such collaboration improves decision-making, especially for tasks requiring deeper reasoning or specialized skills.

Research on LLM-based multi-agent systems exists [6], but there remains a gap in identifying the best technological approaches for building these systems. This survey aims to address this gap by answering two research questions: What state-of-the-art technologies and approaches are available for LLM multi-agent systems? And, which of these are most effective in practice? This survey seeks to identify optimal technologies and methodologies, helping researchers and practitioners navigate the evolving landscape of tools for building advanced multi-agent systems.

\section{Literature Review}
The literature on large language models (LLMs) within multi-agent systems is still emerging, with significant research gaps, particularly in understanding and improving multi-agent paradigms. This review critically examines the current research in multi-agent LLM systems, focusing on architecture, planning, memory, and frameworks. These technological aspects shape the potential and limitations of LLM multi-agent systems, and further investigation can expand their applications and efficiency.

This literature review is structured as given below:

\begin{itemize}
    \item \textbf{Architecture}: Exploring various frameworks like Conquer-and-Merge Discussion (CMD), Chain-of-Agents (CoA), Agent Forest, and Mixture-of-Agents (MoA).
    \item \textbf{Planning}: Discussing frameworks such as AdaPlanner, ChatCoT, KnowAgent, RAP, Tree of Thoughts (ToT), and ReAct.
    \item \textbf{Memory}: Covering the role of memory in LLM systems, with insights on Vector Databases, Retrieval Augmented Generation, ChatDB, MemoryBank, RET-LLM, and Self-Controlled Memory.
    \item \textbf{Technologies / Frameworks}: Examining the technologies and frameworks that facilitate collaboration and task execution, including AutoGen, CAMEL, CrewAI, MetaGPT, and LangGraph.
\end{itemize}

A summary of the findings of each section is provided in Tables I-IV.

\subsection{Architecture}
Research on multi-agent architectures in the LLM space is relatively limited. A significant gap exists in developing frameworks that effectively orchestrate multiple agents to collaborate and solve complex reasoning tasks. While some research has started addressing these challenges, there remains a need for more comprehensive solutions. Below, we analyze some of the key papers exploring architectural designs in multi-agent LLM systems.

\subsubsection{Conquer-and-Merge Discussion (CMD)}
In their paper, Wang et al. [7] present the Conquer-and-Merge Discussion (CMD) framework. This architecture leverages multiple LLM-powered agents that engage in open discussions to address reasoning tasks. Inspired by Minsky's Society of Mind (1988), the CMD framework simulates human-like debates, where each agent contributes different perspectives to improve overall reasoning capabilities.

The CMD framework is structured to allow a group of agents to discuss a question, with each agent generating a viewpoint and explanation in several rounds of interaction. The discussion is guided by a shared history of responses, with agents building on each other's inputs. This design outperforms single-agent methods like the Chain-of-Thought (CoT) approach, as demonstrated in their experiments. However, several limitations persist: the framework simplifies LLM sessions as agents, missing the opportunity to integrate more sophisticated reasoning techniques such as the Tree-of-Thought method or external knowledge bases; CMD has only been tested on reasoning tasks, leaving its applicability to broader domains such as strategic planning or real-time decision-making unexplored; and their experiments were limited to a few LLMs (Bard, Gemini Pro, and ChatGPT-3.5), so further analysis using other models is needed to assess generalizability.

\subsubsection{Chain-of-Agents (CoA)}
Zhang et al. [8], in their paper "Chain of Agents: Large Language Models Collaborating on Long-Context Tasks," propose the Chain-of-Agents (CoA) framework. This architecture is designed for handling long-context tasks that surpass the token limits of individual LLMs. It consists of worker agents that sequentially process portions of the input and pass their results to a manager agent, which aggregates the final output.

The CoA architecture's key innovation is the interleaved read-process method, allowing agents to process chunks of input before receiving the full context. This approach reduces the complexity of handling long-context tasks and enhances interpretability by splitting the work across multiple agents. However, CoA has some limitations: the communication between agents could be improved by leveraging in-context learning or fine-tuning LLMs to optimize their interaction; and the architecture needs further refinement to reduce computational costs and latency, particularly in tasks requiring multiple rounds of communication between agents.

\subsubsection{Agent Forest}
The Agent Forest method, introduced by Li et al. [9] in their paper "More Agents is All You Need," focuses on scaling LLM performance by simply increasing the number of agents. The method employs a sampling-and-voting approach: multiple agents generate responses, and the final answer is determined through majority voting.

This technique demonstrates that increasing the number of agents improves performance, particularly on complex tasks. However, it also reveals limitations: the performance gains from Agent Forest depend on the inherent difficulty of the task, with diminishing returns for overly complex or overly simple tasks; and the approach increases computational costs due to the need for multiple LLM queries, requiring optimization of the sampling phase to enhance cost efficiency.

\subsubsection{Mixture-of-Agents}
The Mixture-of-Agents (MoA) architecture proposed by Wang et al. [10] offers a layered design where agents collaborate in both proposer and aggregator roles. Proposers generate diverse responses, while aggregators synthesize the responses into high-quality outputs. This collaborative model improves LLM performance across multiple benchmarks, including AlpacaEval 2.0 and MT-Bench.

MoA excels by leveraging the strengths of different LLMs, enabling specialized roles for various agents. Its main limitations include the fact that not all LLMs are equally effective in both proposer and aggregator roles, with some models, like WizardLM, excelling as proposers but struggling as aggregators; and although MoA shows impressive results, expanding the number of agents and aggregators introduces complexity in managing collaboration, which may require more sophisticated orchestration techniques in the future.

\begin{table*}[ht]
    \centering
    \caption{Comparison of Architectures in LLM Multi-Agent Systems}
    \begin{tabular}{|p{2.5cm}|p{4cm}|p{4cm}|p{4cm}|}
        \hline
        \textbf{Architecture} & \textbf{Key Features} & \textbf{Strengths} & \textbf{Limitations} \\ 
        \hline
        Conquer-and-Merge Discussion (CMD) & Multiple LLM-powered agents engage in open discussions; Simulates human-like debates to improve reasoning. & Outperforms single-agent methods; Allows agents to build on each other's inputs. & Lacks integration of sophisticated reasoning techniques; Limited testing on broader domains; Analysis required with other models for generalizability. \\ 
        \hline
        Chain-of-Agents (CoA) & Sequential processing of input portions by worker agents; Manager agent aggregates results; Interleaved read-process method. & Reduces complexity of long-context tasks; Enhances interpretability by splitting work across agents. & Communication improvement needed; Further refinement required to reduce computational costs and latency. \\ 
        \hline
        Agent Forest & Employs a sampling-and-voting approach with multiple agents; Performance determined by majority voting. & Performance improves with the number of agents; Effective for complex tasks. & Gains depend on task difficulty; Increased computational costs due to multiple LLM queries. \\ 
        \hline
        Mixture-of-Agents (MoA) & Layered design with proposers and aggregators; Collaborates to generate and synthesize high-quality outputs. & Leverages strengths of different LLMs; Shows impressive results across benchmarks. & Role effectiveness varies among LLMs; Complexity in managing collaboration increases with more agents. \\ 
        \hline
    \end{tabular}
    \label{tab:architectures}
\end{table*}

\subsection{Planning}
In LLM-based multi-agent systems, planning involves the reasoning and action strategies that agents employ to achieve their goals in dynamic environments. These systems must balance reasoning over short and long horizons while responding to environmental feedback. As autonomous decision-makers, LLM agents generate sequences of actions based on initial goals but often require adaptive planning to address the complexity of real-world problems. Effective planning frameworks for LLM multi-agent systems incorporate feedback mechanisms, refining or recalibrating actions based on evolving environmental conditions to avoid issues like hallucination or over-simplification of plans.

\subsubsection{AdaPlanner}
In the AdaPlanner framework, Sun et al. [11] introduce a novel approach that allows LLM agents to refine and adapt plans in response to real-time environmental feedback. This marks a departure from traditional static planning systems, which typically follow a fixed sequence of actions. AdaPlanner’s closed-loop system enables dynamic adjustments, providing critical flexibility for handling complex, long-horizon tasks where static plans often fail due to unexpected changes.

AdaPlanner incorporates two major refinement strategies: in-plan refinement, where agents modify specific parts of an existing plan to address immediate feedback, and out-of-plan refinement, where they create new actions to tackle unforeseen scenarios. To tackle the issue of LLM hallucinations, the authors implement a code-style prompting mechanism, reducing ambiguity in the generated plans and fostering consistency across various tasks. Additionally, AdaPlanner includes a skill discovery feature, allowing agents to reuse successful plans from past tasks as few-shot examples for future problem-solving, effectively improving their adaptability and efficiency.

Despite its advancements, AdaPlanner is not without limitations. Its reliance on few-shot expert demonstrations for more complex tasks remains a constraint, suggesting the need for future research on reducing or eliminating this dependency. Additionally, while its performance in environments like ALFWorld and MiniWoB++ is promising, further testing across diverse domains would better establish the robustness and generalizability of this adaptive planning approach.

\subsubsection{ChatCoT}
Chen et al. [12] introduce ChatCoT, a framework designed to improve large language models (LLMs) in handling complex, multi-step reasoning tasks. Recognizing the limitations of traditional static reasoning models for tasks that require specific knowledge and complex logical steps, ChatCoT employs a tool-augmented chain-of-thought (CoT) reasoning approach tailored to chat-based interactions, such as those in ChatGPT. This framework allows LLMs to alternate between tool manipulation and reasoning actions within a dynamic, multi-turn conversation, enhancing adaptability in complex scenarios.

ChatCoT leverages the strengths of chat-based LLMs by initiating discussions with foundational knowledge about the tools, tasks, and reasoning structure involved. The resulting process is iterative, using step-by-step reasoning that integrates tool manipulation seamlessly with CoT reasoning. Evaluations on datasets such as MATH and HotpotQA show that ChatCoT achieves a 7.9\% relative improvement in performance over existing methods, demonstrating its potential in advancing LLM reasoning capabilities for intricate tasks.

However, ChatCoT has several limitations. The framework has yet to be tested with GPT-4 due to access restrictions, which could affect the generalizability of its findings. Moreover, its design is optimized for chat-based LLMs, which may limit its compatibility with other architectures. Additionally, the high computational requirements, particularly in terms of GPU resources, pose challenges for widespread implementation. Future research will aim to extend ChatCoT's applicability across a broader range of tasks and expand its toolset, potentially enhancing its utility in diverse, complex reasoning scenarios.

\subsubsection{KnowAgent}
In the KnowAgent framework, Zhu et al. [13] present a novel approach to enhance large language models (LLMs) in performing complex reasoning tasks. LLMs, while powerful, often struggle with generating coherent action sequences and interacting effectively with environments due to a lack of inherent action knowledge. KnowAgent addresses this gap by integrating an action knowledge base and employing a self-learning strategy, providing LLMs with structured action knowledge that guides the planning process and mitigates issues like planning hallucinations.

KnowAgent utilizes this action knowledge to refine planning paths, ensuring LLMs generate more reasonable and executable action trajectories. This structured approach translates complex action data into an understandable format for LLMs, significantly enhancing planning accuracy. Experimental results on datasets like HotpotQA and ALFWorld show that KnowAgent not only matches but often exceeds current state-of-the-art performance while effectively reducing planning hallucinations, demonstrating its potential for improving LLM task execution.

Despite its advancements, KnowAgent has limitations. The framework has primarily been evaluated on commonsense question-answering and household tasks, with future potential in domains like medical reasoning, arithmetic, and web browsing. Additionally, KnowAgent currently supports only single-agent applications; exploring multi-agent systems could further enhance its utility through collaborative task execution. Lastly, the manual design of action knowledge bases presents a labor-intensive challenge, suggesting the need for automated solutions to improve adaptability and broaden the framework’s application across diverse environments.

\subsubsection{RAP: Retrieval Augmented Processing}
RAP [14] introduces a groundbreaking framework that enhances the planning capabilities of large language models (LLMs) by integrating retrieval-augmented techniques with contextual memory. As LLMs are increasingly employed as agents for complex decision-making tasks in fields such as robotics, gaming, and API integration, the challenge of incorporating past experiences into current decision-making processes remains significant. To address this, RAP dynamically leverages relevant past experiences tailored to the current context, thereby improving the agents' ability to plan effectively.

What sets RAP apart is its versatility, as it is designed to function seamlessly in both text-only and multimodal environments. This adaptability allows it to tackle a broad spectrum of tasks. Empirical evaluations demonstrate RAP's effectiveness, achieving state-of-the-art performance in textual scenarios and significantly enhancing the capabilities of multimodal LLM agents in embodied tasks. These findings underscore RAP's potential to advance the functionality and applicability of LLM agents in real-world applications that demand sophisticated decision-making.

The RAP framework enables agents to store and retrieve past experiences, guiding subsequent actions based on contextual information extracted from various modalities, including text and images. The results from evaluations across multiple benchmarks reveal that RAP outperforms baseline methods, allowing language agents to flexibly utilize historical experiences in alignment with current situations. This capability mirrors a fundamental human ability, thereby enhancing decision-making capabilities and paving the way for more effective and intelligent LLM-based agents in complex, real-world scenarios.

\subsubsection{Tree of Thoughts (ToT)}
The "Tree of Thoughts" (ToT) framework [15] introduces a novel approach to enhance the problem-solving capabilities of language models (LMs), addressing their limitations in tasks that require exploration, strategic foresight, and the importance of initial decisions. Traditional LMs typically operate within a token-level, left-to-right decision-making paradigm during inference, which can hinder performance in more complex scenarios. ToT expands upon the popular "Chain of Thought" prompting technique, allowing LMs to explore coherent units of text, referred to as "thoughts," that act as intermediate steps in problem-solving.

ToT enables LMs to engage in deliberate decision-making by evaluating multiple reasoning paths and assessing choices, facilitating the ability to look ahead or backtrack as necessary for making more informed global decisions. Experiments demonstrate that ToT significantly improves LMs' performance on three novel tasks that necessitate intricate planning and search: Game of 24, Creative Writing, and Mini Crosswords. For example, in the Game of 24, while GPT-4 with chain-of-thought prompting achieved only a 4\% success rate, the ToT framework elevated this figure to 74\%.

In the limitations and future directions section, the authors note that deliberate search methods like ToT may not be essential for many tasks where GPT-4 already performs well. The initial exploration is limited to three relatively simple tasks, indicating a need for further research into more complex decision-making applications, such as coding, data analysis, and robotics. Additionally, the resource-intensive nature of search methods compared to sampling methods may pose challenges, though the modular flexibility of ToT allows users to tailor performance-cost tradeoffs. Ongoing open-source initiatives could further reduce associated costs. The potential for fine-tuning LMs using ToT-style high-level counterfactual decision-making is also highlighted as an avenue for improving LMs' problem-solving abilities.

\subsubsection{ReAct}
The "ReAct" framework [16] represents a significant advancement in the integration of reasoning and acting within large language models (LLMs). While LLMs have showcased remarkable capabilities in language understanding and interactive decision-making, the traditional separation of reasoning (e.g., chain-of-thought prompting) and acting (e.g., action plan generation) has limited their effectiveness in complex tasks. ReAct addresses this limitation by facilitating an interleaved generation of reasoning traces and task-specific actions, enhancing synergy between the two processes. This approach allows reasoning traces to aid the model in inducing, tracking, and updating action plans while also managing exceptions, whereas actions enable interaction with external sources, such as knowledge bases or environments.

The application of ReAct across various language and decision-making tasks demonstrates its superiority over state-of-the-art baselines, providing enhanced human interpretability and trustworthiness. In experiments involving question answering (HotpotQA) and fact verification (Fever), ReAct effectively mitigates common issues of hallucination and error propagation found in chain-of-thought reasoning by leveraging a simple Wikipedia API. This interaction leads to more human-like task-solving trajectories that are more interpretable than those produced by methods lacking reasoning traces.

ReAct introduces a straightforward yet powerful method for synergizing reasoning and acting within LLMs, yielding superior performance and interpretable decision traces across diverse tasks, including multi-hop question answering, fact-checking, and interactive decision-making. Given the positive results achieved, this approach is highly recommended for enhancing LLMs' action planning capabilities. Although the simplicity of ReAct presents advantages, it also reveals the need for more demonstrations to effectively learn complex tasks with large action spaces. Initial experiments indicate that fine-tuning on specific tasks, such as HotpotQA, may further improve performance, particularly through the incorporation of high-quality human annotations. Exploring multi-task training and integrating ReAct with complementary paradigms like reinforcement learning could lead to the development of more robust agents, further unlocking the potential of LLMs for diverse applications.

\begin{table*}[ht]
    \centering
    \caption{Comparison of Planning Frameworks in LLM Multi-Agent Systems}
    \begin{tabular}{|p{2.5cm}|p{4cm}|p{4cm}|p{4cm}|}
        \hline
        \textbf{Framework} & \textbf{Key Features} & \textbf{Strengths} & \textbf{Limitations} \\ 
        \hline
        AdaPlanner & Dynamic adjustments to plans based on real-time feedback; In-plan and out-of-plan refinement strategies; Code-style prompting for reducing ambiguity. & Flexibility in complex, long-horizon tasks; Improved adaptability and efficiency through skill discovery. & Relies on few-shot expert demonstrations; Performance needs further testing across diverse domains. \\ 
        \hline
        ChatCoT & Tool-augmented chain-of-thought reasoning for chat-based interactions; Iterative, multi-turn conversations. & Achieves 7.9\% relative improvement in performance on complex tasks; Enhances adaptability in reasoning scenarios. & Limited testing with GPT-4; High computational requirements; Optimized for chat-based models. \\ 
        \hline
        KnowAgent & Integration of an action knowledge base with self-learning strategies; Refines planning paths for coherent action sequences. & Matches or exceeds state-of-the-art performance; Reduces planning hallucinations effectively. & Primarily evaluated on specific tasks; Currently supports only single-agent applications; Manual design of action knowledge bases is labor-intensive. \\ 
        \hline
        RAP & Retrieval-augmented techniques integrated with contextual memory; Functions in both text-only and multimodal environments. & Achieves state-of-the-art performance in various benchmarks; Enhances decision-making by utilizing past experiences. & Adapting for multimodal tasks requires careful implementation; Complexity in managing retrieval processes. \\ 
        \hline
        {ReAct} & Interleaves reasoning and acting; Facilitates decision traces and task-specific actions; Enhances interpretability and trustworthiness. & Superior performance across diverse tasks; Effective in mitigating hallucination and error propagation; Highly recommended for action planning capabilities. & Simplicity may limit complexity handling; Further demonstrations needed for large action spaces. \\ 
        \hline
    \end{tabular}
    \label{tab:planning_frameworks}
\end{table*}

\subsection{Memory}
Memory plays a critical role in enhancing the capabilities of agents to retain and retrieve information relevant to their tasks. Effective memory systems enable agents to recall past interactions and experiences, facilitating informed decision-making and improving task performance in dynamic environments. The integration of advanced memory mechanisms, such as vector databases and retrieval-augmented generation, allows agents to store vast amounts of information while maintaining quick access to pertinent data. Additionally, these memory systems can implement self-controlled mechanisms to ensure that agents prioritize the most relevant memories, enabling them to adapt to new challenges while avoiding the pitfalls of irrelevant or outdated information. By enhancing memory functionality, LLM agents can better navigate complex tasks and interactions, leading to more robust and intelligent behavior.

\subsubsection{Vector Databases}
Jing et al. [17] provide a comprehensive survey of the intersection between large language models (LLMs) and vector databases (VecDBs), a rapidly evolving area of research aimed at addressing critical limitations in LLM-based systems. Despite the impressive capabilities of LLMs, they struggle with issues such as hallucinations, memory constraints, outdated knowledge, and the high costs associated with commercial deployment. VecDBs offer a promising solution to these challenges by efficiently storing and retrieving the high-dimensional vector representations that are fundamental to LLM operations.

The integration of LLMs and VecDBs enhances the ability of LLM systems to manage and retrieve vast amounts of information, reducing reliance on static memory and enabling more dynamic, context-aware interactions. By leveraging VecDBs, LLMs can access external knowledge bases, mitigating hallucinations and outdated information while improving response accuracy. Additionally, this synergy helps overcome memory limitations by enabling the offloading of knowledge, allowing for scalable, long-term storage solutions. This framework has paved the way for the development of memory systems specifically designed for LLMs, advancing their ability to manage complex tasks over extended interactions.

The paper highlights both the opportunities and challenges in combining LLMs and VecDBs, categorizing existing research into distinct prototypes and interdisciplinary approaches. It also addresses the engineering challenges related to optimizing this integration, such as designing efficient data retrieval mechanisms and ensuring compatibility with LLM architectures. Looking ahead, the authors call for further research into expanding the utility of VecDBs in diverse LLM applications, driving advancements in data handling, knowledge extraction, and the development of robust memory solutions for LLM-based systems.

\subsubsection{Retrieval Augmented Generation}
Lewis et al. [18] introduce the Retrieval-Augmented Generation (RAG) framework, a novel approach aimed at enhancing the performance of large pre-trained language models on knowledge-intensive tasks. Traditional LLMs, while capable of storing vast amounts of factual knowledge within their parameters, often struggle with accessing and manipulating this knowledge in a precise and scalable way. Furthermore, updating their knowledge or providing provenance for decisions remains a significant challenge. RAG addresses these limitations by combining parametric memory, based on a pre-trained sequence-to-sequence (seq2seq) model, with non-parametric memory, represented as a dense vector index of external sources like Wikipedia, accessed via a neural retriever.

By integrating these two memory systems, RAG allows models to retrieve relevant information dynamically from external sources during language generation, rather than solely relying on the knowledge embedded in their parameters. This dual-memory design significantly improves the specificity, diversity, and factual accuracy of generated responses, particularly in tasks such as open-domain question answering (QA) and knowledge-intensive text generation. The authors present two variants of RAG: one that retrieves the same passage for the entire sequence and another that retrieves different passages for each token, further refining the generation process. Through extensive experiments, RAG outperforms state-of-the-art parametric seq2seq models and retrieve-and-extract architectures, setting new benchmarks in multiple QA tasks.

RAG’s impact goes beyond immediate performance gains; it laid the groundwork for future developments in LLM memory systems. The ability to hot-swap the retrieval index without retraining the model provides a scalable solution for updating LLMs with new knowledge, addressing a critical issue in long-term LLM use. This hybrid memory structure has inspired subsequent memory-augmented solutions, paving the way for more sophisticated LLMs capable of seamlessly integrating parametric and non-parametric knowledge sources to enhance reasoning, generation, and decision-making across a wide range of tasks.

\subsubsection{ChatDB}
In their paper, Hu et al. [19] introduce a novel framework that enhances large language models (LLMs) by integrating symbolic memory, represented by SQL databases. The motivation stems from the limitations of neural memory mechanisms, which are prone to error accumulation and struggle with complex reasoning tasks. By incorporating symbolic memory, ChatDB enables more precise and reliable memory manipulation, allowing LLMs to perform multi-hop reasoning through interaction with an external database. This is inspired by modern computer architectures rather than biological models, providing a more robust solution for advanced reasoning tasks.

The ChatDB framework operates in three main stages. First, in the input processing stage, the system generates SQL instructions to interact with the database if memory is required; otherwise, the LLM responds directly. In the chain-of-memory stage, a series of SQL operations such as select, update, insert, and delete are executed, with each step influencing the next based on the results of previous operations. Finally, in the response summary stage, ChatDB generates a coherent final output based on the results obtained from manipulating the symbolic memory, ensuring accurate and logical responses.

The experimental results show that ChatDB outperforms models like ChatGPT, especially in tasks requiring complex reasoning, by eliminating error propagation through precise memory operations. By using SQL as a symbolic memory language, ChatDB introduces a reliable method for LLMs to handle intermediate results, enhancing both the accuracy and capability of the model in various management and reasoning scenarios. This symbolic memory approach sets the stage for further advancements in memory-augmented LLMs, offering a more scalable and efficient framework for handling knowledge-intensive tasks.

\subsubsection{MemoryBank}
Zhong et al. [20] introduce MemoryBank, an innovative memory mechanism designed to address a significant limitation in large language models (LLMs): the lack of long-term memory. While LLMs have made remarkable strides in performing various tasks, their inability to maintain and recall information from past interactions has hindered their performance in applications requiring sustained context, such as personal assistance or therapy. MemoryBank enables LLMs to store, retrieve, and update memories dynamically, allowing models to evolve in understanding users’ personalities over time. By integrating a memory updater based on the Ebbinghaus Forgetting Curve theory, the system can selectively forget or reinforce memories depending on their relevance and the time elapsed, mirroring human memory retention patterns.

MemoryBank operates around three key components: a memory storage system for data retention, a memory retriever to summon context-specific memories, and a memory updater inspired by psychological principles. This updater ensures that the system adapts over time, retaining essential information while allowing less significant memories to fade. This anthropomorphic memory mechanism enhances user interactions, providing more personalized responses and a deeper understanding of user behavior. The framework is versatile, functioning across both closed-source models, such as ChatGPT, and open-source models, including ChatGLM. MemoryBank demonstrates its capabilities through the chatbot SiliconFriend, designed for long-term companionship, which uses MemoryBank to recall past conversations and adjust its responses based on user preferences and emotional state.

MemoryBank significantly improves the ability of LLMs to handle long-term interactions by offering a scalable solution for memory retention and recall. SiliconFriend, equipped with this mechanism, demonstrates the potential for AI systems to deliver more empathetic and personalized experiences. MemoryBank’s flexible structure and memory updating mechanism allow LLMs to provide relevant and accurate information across extended dialogues, setting the stage for further advancements in memory-augmented LLMs. This framework not only enhances LLM performance in personal companion systems but also lays the groundwork for future developments in AI-human interaction, where long-term memory plays a critical role in delivering meaningful and sustained engagements.

\subsubsection{RET-LLM}
Modarressi et al. [21] present RET-LLM, a groundbreaking framework designed to enhance large language models (LLMs) by integrating a general read-write memory unit. Despite the remarkable advancements LLMs have made in natural language processing (NLP), their lack of a dedicated memory system restricts their ability to store and retrieve knowledge explicitly for diverse tasks. RET-LLM addresses this gap by allowing LLMs to extract, store, and recall information as needed, improving their task performance. Drawing inspiration from Davidsonian semantics theory, the memory unit captures knowledge in the form of triplets, facilitating a scalable and interpretable memory structure that can be easily updated and aggregated.

The RET-LLM architecture comprises three key components: a Controller, a Fine-tuned LLM, and a Memory unit. The Controller regulates the information flow among the user, the LLM, and the Memory unit, ensuring efficient communication. The Fine-tuned LLM processes incoming text and determines when to invoke memory. To facilitate memory interaction, the framework implements a text-based API schema, allowing the LLM to generate standardized memory API calls. Knowledge is stored in a triplet format, structured as <first argument, relation, second argument>, reflecting the theoretical principles of Davidsonian semantics. This organization enables effective management of relational knowledge, allowing the model to perform better in various NLP tasks, particularly in question answering.

RET-LLM significantly enhances LLM capabilities by enabling the explicit storage and retrieval of information, thereby addressing one of the critical limitations of traditional LLMs. The framework's triplet-based memory structure allows for nuanced relationships to be stored and accessed, showcasing superior performance in question answering tasks, especially those requiring temporal reasoning. Preliminary qualitative evaluations indicate that RET-LLM outperforms baseline approaches, demonstrating its potential in effectively managing time-dependent information. Although still under development, future iterations of RET-LLM will focus on comprehensive empirical evaluations using real datasets and refining the fine-tuning process to broaden its applicability across various types of informative relations. The ongoing research highlights the transformative potential of incorporating a robust memory unit into LLMs, paving the way for more intelligent and adaptable AI systems.

\subsubsection{Self-Controlled Memory}
Wang et al. [22] introduce the Self-Controlled Memory (SCM) framework, a novel approach aimed at enhancing large language models (LLMs) by addressing their limitations in processing lengthy inputs. Traditional LLMs often struggle to retain critical historical information, which hinders their performance in tasks requiring long-term memory. The SCM framework consists of three main components: an LLM-based agent that serves as the backbone of the system, a memory stream that stores agent memories, and a memory controller that updates these memories and determines when and how to utilize them. Importantly, SCM operates in a plug-and-play manner, enabling seamless integration with any instruction-following LLMs without the need for extensive modifications or fine-tuning.

To validate the effectiveness of the SCM framework, the authors annotated a dataset designed for evaluating its capabilities in handling ultra-long texts across three tasks: long-term dialogues, book summarization, and meeting summarization. Experimental results reveal that the SCM framework significantly improves retrieval recall and generates more informative responses compared to competitive baselines in long-term dialogue scenarios. These findings demonstrate SCM's potential to enhance the performance of LLMs, allowing them to better manage extensive conversations and detailed summarizations, ultimately addressing a key challenge in the field of natural language processing.

Despite its advantages, the SCM framework has some limitations, particularly regarding the evaluation of its performance in infinite dialogue settings, which were tested only up to 200 dialogue turns and a maximum token count of 34,000. This constraint arises from the challenges associated with qualitatively and quantitatively evaluating very long texts. Additionally, the effectiveness of the SCM framework relies on powerful instruction-following LLMs like text-davinci-003 and gpt-3.5-turbo-0301. However, the authors anticipate that future advancements in smaller, more powerful LLMs could mitigate this limitation. Overall, the SCM framework offers a promising direction for extending the input length of LLMs and improving their ability to capture and recall useful information from historical data.

\begin{table*}[ht]
    \centering
    \caption{Comparison of Memory Frameworks in LLM Multi-Agent Systems}
    \begin{tabular}{|p{2.5cm}|p{4cm}|p{4cm}|p{4cm}|}
        \hline
        \textbf{Framework} & \textbf{Key Features} & \textbf{Strengths} & \textbf{Limitations} \\ 
        \hline
        Vector Databases (VecDB) & Efficient storage and retrieval of high-dimensional vector representations; Enhances LLMs’ ability to access external knowledge bases. & Reduces reliance on static memory; Mitigates hallucinations and outdated information; Enables dynamic, context-aware interactions. & Integration challenges; Requires optimization for compatibility with LLM architectures. \\ 
        \hline
        Retrieval-Augmented Generation (RAG) & Combines parametric and non-parametric memory; Retrieves relevant information during language generation. & Improves specificity, diversity, and factual accuracy; Scalable solution for updating LLMs with new knowledge. & Complexity in retrieval index management; Requires extensive training data for optimal performance. \\ 
        \hline
        ChatDB & Integrates symbolic memory via SQL databases; Supports multi-hop reasoning through external database interaction. & Enables precise memory manipulation; Reduces error propagation; Enhances performance in complex reasoning tasks. & Limited to tasks that can be mapped to SQL operations; Requires careful SQL instruction generation. \\ 
        \hline
        MemoryBank & Offers long-term memory retention and dynamic updating; Uses the Ebbinghaus Forgetting Curve for selective memory management. & Enhances personalized responses; Adapts over time to user interactions; Versatile for various LLM architectures. & Performance may vary based on user interaction context; Requires careful memory updating to avoid overload. \\ 
        \hline
        RET-LLM & Integrates a read-write memory unit; Stores knowledge as triplets for scalable, interpretable memory management. & Addresses LLM limitations in storing and retrieving knowledge; Superior performance in question answering tasks. & Still under development; Requires comprehensive evaluations and real dataset testing. \\ 
        \hline
        Self-Controlled Memory (SCM) & Consists of an agent, memory stream, and memory controller; Enhances handling of lengthy inputs. & Significantly improves retrieval recall; Generates informative responses; Easy integration with existing LLMs. & Limited evaluation in infinite dialogue settings; Dependent on powerful instruction-following LLMs for effectiveness. \\ 
        \hline
    \end{tabular}
    \label{tab:memory_frameworks}
\end{table*}

\subsection{Technologies / Frameworks}
The development of LLM-based multi-agent systems relies heavily on a variety of technologies and frameworks that facilitate efficient agent collaboration and task execution. These frameworks provide essential tools for building, deploying, and managing multi-agent environments, enabling seamless communication and coordination among agents. Technologies such as AutoGen and MetaGPT empower agents to generate dynamic responses and solutions based on real-time data and interactions. Additionally, frameworks like CAMEL and CrewAI offer integrated environments that streamline the design and orchestration of multi-agent systems, allowing for enhanced scalability and flexibility. By leveraging these technologies, researchers and practitioners can create sophisticated LLM-based systems that adapt to changing circumstances and optimize their performance across a range of applications, from robotics to intelligent assistance.

\subsubsection{AutoGen}
Wu et al. [23] present AutoGen, an open-source framework designed to facilitate the development of large language model (LLM) applications through multi-agent conversations. This innovative framework allows multiple agents to interact, collaborate, and accomplish tasks by leveraging customizable and conversable agents that can operate in diverse modes. AutoGen supports a combination of LLMs, human inputs, and various tools, enabling developers to flexibly define agent interaction behaviors. By employing both natural language and computer code for programming conversation patterns, AutoGen serves as a generic framework that caters to a wide range of applications, including mathematics, coding, question answering, operations research, and online decision-making.

To streamline the creation of complex LLM applications, AutoGen is built upon the principles of conversable agents and conversation programming. A conversable agent can send and receive messages to engage with other agents while maintaining its internal context. This modular approach allows for a variety of capabilities, powered by LLMs, tools, or human input. Conversation programming encompasses two key concepts: computation, which pertains to the actions agents undertake in a multi-agent conversation, and control flow, which dictates the sequence or conditions under which these actions occur. This conversation-centric paradigm simplifies the reasoning behind complex workflows, allowing agents to pass messages dynamically and adaptively as they collaborate.

The authors emphasize that AutoGen enhances multi-agent cooperation through its unified conversation interface and auto-reply mechanisms, effectively harnessing the strengths of chat-optimized LLMs. The framework enables developers to create and experiment with multi-agent systems that can be reused, customized, and extended, all while significantly reducing development effort. Experimental results indicate that AutoGen outperforms state-of-the-art approaches, streamlining the development process and enabling flexible, dynamic interactions among agents. While still in the early stages of development, AutoGen lays the groundwork for future research into the integration of existing agent implementations, optimal agent topologies, and the balance between automation and human control in multi-agent workflows, addressing potential safety challenges as the complexity of applications increases.

\subsubsection{CAMEL}
Li et al. [24] present CAMEL, a novel framework aimed at enhancing the autonomous cooperation of communicative agents in chat-based language models. As these models continue to evolve, their effectiveness often hinges on human input to guide conversations, which can be a daunting and time-consuming task. The authors propose a role-playing approach that employs inception prompting to enable agents to work collaboratively toward task completion while staying aligned with human intentions. This framework not only facilitates the generation of conversational data but also serves as a valuable resource for exploring the behaviors and capabilities of a society of agents, particularly in multi-agent settings focused on instruction-following cooperation.

The paper highlights the significance of autonomous cooperation among communicative agents and delineates the challenges that accompany it, such as conversation deviation, role flipping, and defining termination conditions. The role-playing framework offers a scalable solution to these challenges, allowing agents to engage in effective collaboration with minimal human intervention. The authors conducted comprehensive evaluations to assess the framework's effectiveness, demonstrating that it leads to better outcomes in task completion. Additionally, their open-sourced library includes implementations of various agents, data generation pipelines, and analytical tools, thus fostering research on communicative agents and advancing the understanding of cooperative behaviors in multi-agent systems.

By providing insights into the complexities of agent interactions and the dynamics of cooperative AI systems, this work significantly contributes to the growing field of large language models. The framework not only emphasizes the potential for autonomous agent collaboration but also sets the stage for future research endeavors aimed at improving the scalability and efficacy of communicative agents in diverse applications. With CAMEL, Li et al. pave the way for more sophisticated interactions among agents, enhancing the capabilities of language models and their applications in real-world scenarios.

\subsubsection{CrewAI}
In the paper by Berti et al. [25], the CrewAI framework is introduced as a crucial component for implementing the AI-Based Agents Workflow (AgWf) paradigm, aimed at enhancing process mining (PM) tasks through the integration of Large Language Models (LLMs). CrewAI serves as a Python framework that facilitates the design and execution of AgWf, enabling developers to harness the capabilities of LLMs in a structured manner. The framework is built upon several key concepts: AI-based agents, AI-based tasks, and tools. AI-based agents combine LLMs with tailored system prompts, effectively aligning the model’s behavior with specific roles. This role prompting is essential for ensuring that the agents perform their designated tasks accurately.

Within the CrewAI framework, AI-based tasks are defined through textual instructions linked to these AI-based agents, allowing for a clear delineation of responsibilities. Furthermore, tools are implemented as Python classes or functions, which can be integrated into tasks based on their documentation strings, including input parameters and output types. The framework supports both traditional sequential execution of tasks and more complex concurrent execution through hierarchical processes, although further development is needed in this area. By decomposing complex PM tasks into simpler, manageable workflows, CrewAI aims to enhance the reasoning capabilities of LLMs, thus addressing the limitations that arise when these models are faced with intricate scenarios.

The CrewAI framework exemplifies a modern approach to leveraging AI for process mining by combining the strengths of LLMs with deterministic tools to produce high-quality outputs. The paper details various AI-based tasks that can be employed within CrewAI for PM applications, including prompt optimizers, ensembles, routers, evaluations, and output improvers. Through practical examples such as root cause analysis and bias detection in process mining event logs, Berti et al. demonstrate the potential of CrewAI to revolutionize how process mining tasks are approached in the era of AI-based agents. The framework not only provides a pathway for implementing effective workflows but also encourages further research into automating workflow definitions and enhancing agent evaluation frameworks.

\subsubsection{MetaGPT}
The paper by Hong et al. [26] introduces MetaGPT, an innovative meta-programming framework designed to enhance collaboration among multi-agent systems built on large language models (LLMs). Existing LLM-based multi-agent systems excel at simple dialogue tasks but struggle with complex scenarios due to logic inconsistencies and cascading hallucinations that arise from naively chaining LLMs together. MetaGPT addresses these challenges by incorporating efficient human workflows through the encoding of Standardized Operating Procedures (SOPs) into structured prompt sequences. This approach allows agents with human-like domain expertise to verify intermediate results, thus reducing errors and improving overall performance.

A key feature of MetaGPT is its assembly line paradigm, which efficiently assigns diverse roles to various agents, breaking down complex tasks into manageable subtasks that promote effective collaboration. The framework emphasizes role specialization and structured communication, enhancing the agents' ability to interact and share information. By implementing a communication protocol that includes structured interfaces and a publish-subscribe mechanism, agents can access relevant information from other roles and the environment, thereby streamlining the workflow and facilitating a more coherent solution generation process.

MetaGPT represents a significant advancement in the development of LLM-based multi-agent systems, combining flexibility and convenience with robust functionality. The integration of human-like SOPs within the framework minimizes unproductive collaboration, while the novel executable feedback mechanism allows for real-time debugging and code execution during runtime, leading to notable improvements in code generation quality. MetaGPT's impressive performance on benchmarks like HumanEval and MBPP underscores its potential as a valuable tool for future research and application in multi-agent collaborations, paving the way for more effective and coherent solutions in complex problem-solving scenarios.

\subsubsection{LangGraph}
The LangGraph framework [27] emerges as a powerful tool for developing advanced Retrieval-Augmented Generation (RAG) systems, particularly for knowledge-based question-answering (QA) applications. Unlike traditional RAG models that often suffer from accuracy degradation due to their reliance on static pre-loaded knowledge, LangGraph leverages graph technology to enhance the information retrieval process. By enabling efficient searches and evaluations of the reliability of retrieved data, LangGraph significantly improves the contextual understanding and accuracy of generated responses. This innovative approach not only mitigates the limitations of existing RAG models but also facilitates the integration of real-time data, allowing for a more dynamic and accurate information synthesis process.

LangGraph stands out among other frameworks by providing a stateful, multi-actor application environment specifically designed for LLMs. Its capability to create agent workflows as cyclic graph structures allows developers to define intricate flows and control the state of the application, which is essential for building reliable agents. The LangGraph Conversational Retrieval Agent further enhances this by incorporating language processing, AI model integration, and graph-based data management, making it an ideal option for crafting sophisticated language-based AI applications. Its architecture encourages collaborative interactions among agents, ensuring that complex tasks are handled with precision and reliability.

Overall, the implementation of the LangGraph framework within the context of advanced RAG systems offers a compelling advantage over previously mentioned frameworks. Its focus on creating cyclic workflows not only allows for a more robust and efficient handling of multi-agent tasks but also significantly elevates the quality of responses through improved data processing and reliability assessment. The framework’s ability to enhance real-time data accessibility and support diverse question types positions it as an invaluable resource for developing high-quality generative AI services, particularly in customer support and information retrieval applications. As demonstrated in Jeong’s study, LangGraph provides a crucial foundation for advancing the capabilities of RAG-based systems, making it a preferred choice for researchers and developers in the field.

\begin{table*}[ht]
    \centering
    \caption{Comparison of Technologies and Frameworks in LLM Multi-Agent Systems}
    \begin{tabular}{|p{2.5cm}|p{4cm}|p{4cm}|p{4cm}|}
        \hline
        \textbf{Framework} & \textbf{Key Features} & \textbf{Strengths} & \textbf{Limitations} \\ 
        \hline
        AutoGen & Open-source framework for multi-agent conversations; Supports diverse modes of agent interaction; Combines LLMs, human inputs, and tools. & Enhances multi-agent cooperation; Flexible definition of agent behaviors; Outperforms state-of-the-art approaches. & Still in early development; Integration of existing agent implementations requires further research. \\ 
        \hline
        CAMEL & Role-playing approach with inception prompting; Facilitates task completion among communicative agents; Generates conversational data. & Autonomous cooperation reduces human input; Scalable solution to conversation challenges; Includes open-sourced agent implementations. & Challenges with conversation deviation and role flipping; Requires evaluation in diverse contexts. \\ 
        \hline
        CrewAI & Framework for AI-Based Agents Workflow (AgWf); Integrates LLMs with AI-based tasks and tools; Supports sequential and concurrent task execution. & Enhances reasoning capabilities of LLMs; Breaks down complex tasks into manageable workflows; Encourages high-quality outputs. & Further development needed for concurrent execution; Manual design of agents can be complex. \\ 
        \hline
        MetaGPT & Meta-programming framework that integrates Standardized Operating Procedures (SOPs); Assigns roles to agents and enhances structured communication. & Reduces errors through human-like verification; Improves collaboration and solution generation; Notable performance improvements in benchmarks. & Complexity in implementation; Reliance on structured prompts may limit flexibility. \\ 
        \hline
        LangGraph & Enhances Retrieval-Augmented Generation (RAG) systems with graph technology; Allows cyclic workflows for agent applications. & Improves accuracy and contextual understanding; Enables real-time data integration; Supports collaborative interactions among agents. & Complexity in managing cyclic workflows; Reliance on graph structures may pose a learning curve. \\ 
        \hline
    \end{tabular}
    \label{tab:technologies_frameworks}
\end{table*}

\section{Methodology}
This review aimed to systematically evaluate and synthesize existing research on large language model (LLM) multi-agent systems, specifically addressing the aspects that directly support their application and scalability, while ensuring that a methodology synonymous with the standard practice of following the scientific method was utilized.

To focus the scope effectively, research questions were first defined to limit the exploration to technologies explicitly designed for LLM multi-agent systems rather than those pertaining to LLMs and multi-agent systems independently. This choice allowed for a thorough examination of the unique intersections between LLMs and multi-agent interactions, avoiding the dilution of findings across broader, less targeted studies.

To cover the breadth of critical topics in the field, four primary aspects were identified: Architecture, Memory, Planning, and Technologies/Frameworks. Each aspect addresses an essential component in the design and operation of LLM-based multi-agent systems, reflecting the distinctive requirements and challenges of these systems.

The literature search was conducted across multiple well-regarded academic sources, including Google Scholar, IEEE Xplore, and arXiv, using targeted keywords associated with each of the four topics. Among these, arXiv proved to be the most valuable repository, providing a high concentration of relevant papers that detailed recent developments and experimental applications of LLM-based multi-agent systems. Each paper was evaluated for its relevance to the identified topics, with priority given to publications from credible authors and reputable conferences or journals. This selection process ensured that the reviewed literature included the most influential and innovative work within the field.

For each shortlisted paper, detailed content analysis was performed, with particular attention to descriptions of methods, architectures, and experimental setups. The merits and limitations of each approach were recorded, allowing for a comprehensive understanding of current capabilities, typical challenges, and areas with potential for improvement. This analysis informed recommendations for researchers and engineers in the field, offering guidance on optimal practices and common pitfalls in developing LLM-based multi-agent systems.

The findings from this review are synthesized to highlight prominent trends and identify future research opportunities, such as advancements in scalability and robustness. The focus on these emerging needs aims to guide ongoing research efforts in building systems that can effectively manage complex multi-agent interactions. An overview of this methodology is depicted in the following diagram.

\begin{figure}[ht]
    \centering
    \includegraphics[width=0.6\linewidth]{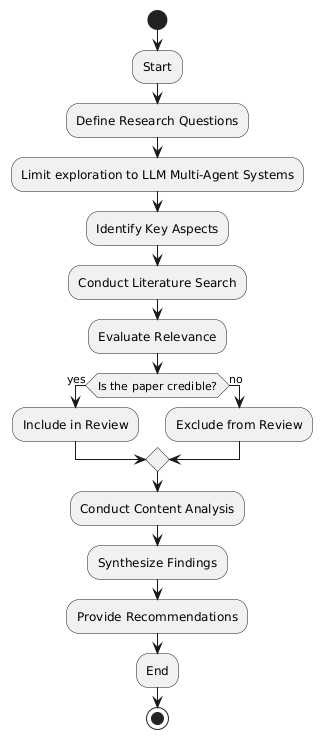}
    \caption{Overview of Survey Methodology}
    \label{fig:your_label}
\end{figure}

\section{Discussion}

\subsection{Key Findings}
Upon analyzing various architectural approaches for LLM-based multi-agent systems, the "Mixture of Agents" (MoA) architecture proposed by Wang et al. emerges as a highly effective design for achieving sophisticated collaboration among agents. MoA’s layered model differentiates agents into proposer and aggregator roles, with proposers generating diverse responses and aggregators synthesizing them into cohesive, high-quality outputs. This architecture enhances LLM performance across benchmarks like AlpacaEval 2.0 and MT-Bench, underscoring its versatility. By allowing specialized roles for different agents, MoA maximizes the strengths of various LLMs. However, MoA does have limitations, as not all LLMs function equally well in both roles; for instance, WizardLM excels as a proposer but struggles in the aggregator capacity. Furthermore, scaling MoA with additional agents introduces complexity, suggesting that future implementations may benefit from advanced orchestration techniques to manage the increasing number of interactions effectively.

Regarding memory in LLM multi-agent systems, the analysis found that various memory approaches could be equally applicable, depending on the specific use case. Short-term memory models, for example, excel in scenarios where agents need rapid access to recent information but do not necessarily require extensive historical context. Conversely, long-term memory models are valuable for applications that demand more in-depth information retention over extended interactions. The choice of memory architecture should align with the intended function of the system, as this can have a significant impact on performance and scalability, especially in cases that require high responsiveness or nuanced historical recall.

For planning, the ReAct framework stands out as a preferred approach for integrating reasoning and action planning within LLM-based multi-agent systems. By enabling an interleaved generation of reasoning traces and task-specific actions, ReAct effectively addresses traditional limitations in complex task handling. This framework synergizes reasoning and action, allowing the model to update and manage plans while interacting with external sources like knowledge bases. ReAct has demonstrated its strengths across diverse tasks, including multi-hop question answering and fact verification, by mitigating common issues such as hallucination and error propagation. While ReAct provides an elegant solution for task planning, it could benefit from enhancements in action space handling, with possibilities for multi-task training or integration with reinforcement learning to expand its robustness and applicability.

In terms of technologies and frameworks for developing LLM-based multi-agent systems, this review found that the choice of framework often depends on the requirements of the application rather than any inherent superiority of one framework over another. Factors such as ease of integration, support for specific programming languages, scalability, and compatibility with external data sources play a significant role in determining which framework is optimal. As the field continues to evolve, the adaptability of frameworks to new advances and compatibility with emerging tools for LLMs will be crucial for their sustained utility.

\subsection{Future Directions}
In summary, this review has identified the Mixture of Agents architecture and ReAct planning framework as highly effective strategies for designing and managing LLM-based multi-agent systems. Memory and technology choices remain largely application-specific, underscoring the importance of aligning system components with the desired outcomes. Together, these findings present a roadmap for future developments in LLM multi-agent systems, where ongoing research and refined frameworks will likely contribute to increasingly robust, versatile applications in this domain.

\section{Conclusion}
The development of LLM-based multi-agent systems marks a significant step forward in enabling complex, collaborative AI applications. This review has identified core frameworks and methodologies that enhance system effectiveness, such as the Mixture of Agents (MoA) for structured agent collaboration and the ReAct framework for integrating reasoning and action. Memory architectures within multi-agent systems remain diverse, with application-specific requirements determining the most suitable approach for short-term or long-term data retention. Technology choices hinge on application demands, favoring adaptable frameworks capable of integrating external data and supporting scalability. Although these systems exhibit great potential, ongoing challenges—including computational costs, communication optimization, and role specialization—must be addressed for broader applicability. Future research and practical advancements are essential to evolving these frameworks, promoting resilient and flexible LLM-based multi-agent systems poised to support sophisticated AI-driven workflows in varied domains.

\end{document}